\documentstyle{aipproc}
\begin{document}



\font\xbit=cmmib10
\def\bolditalx{\fam 0 \xbit}

\font\xiibit=cmmib12
\def\boldital{\fam 0 \xiibit}

\font\viiibit=cmmib8
\def\ssboldital{\fam 0 \viiibit}

\font\viiisb=cmbsy8
\def\ssboldsym{\fam 0 \viiisb}

\font\tenbb=msbm10
\def\bbten{\fam 0 \tenbb}

\font\sevenbb=msbm7
\def\bbsub{\fam 0 \sevenbb}

\font\teneu=eufm10
\def\euler{\fam 0 \teneu}


\def \ns{\enspace}
\def \ts{\thinspace}
\def \nts{\negthinspace}
\def \beq{\begin{equation}}
\def \eeq{\end{equation}}
\def \beqa{\begin{eqnarray}}
\def \eeqa{\end{eqnarray}}
 

\def \LQCD{\Lambda_{\rm QCD}}
\def \xf{x_{{}_F}}
\def \mn{m}
\def \CHI{{\cal X}}
\def \curlyD{{\cal D}}
\def \curlyL{{\cal L}}
\def \DIII{{\cal D}}
\def \qt{\hbox{\boldital q}}
\def \qqq{\hbox{\bolditalx q}}
\def \xt{\hbox{\boldital x}}
\def \deltat{{\bf \Delta}}
\def \qte{\hbox{\ssboldital q}}
\def \xte{\hbox{\ssboldital x}}
\def \deltate{{\bf \Delta}}
\def \gnumX{ {{dN}\over{d\xf d^2\qt}} }


\title{From Cr\^epes to Pancakes\\
 in the MV Model\footnote{Talk 
presented at the 7th Conference on the Intersections of Particle 
and Nuclear Physics, Qu\'ebec City, Qu\'ebec, Canada, 
May 22--28, 2000.}}
\author{Gregory Mahlon}
\address{Department of Physics, McGill University\\
3600 University Street, Montr\'eal, Qu\'ebec H3A 2T8, Canada}


\maketitle

\begin{abstract}
The McLerran-Venugopalan model provides a framework which allows 
one to compute the gluon distribution function of a very large nucleus
from the equations of QCD, provided that the longitudinal momentum 
fraction, $\xf$, is sufficiently small. The source of color charge 
for this computation may be thought of as a cr\^epe moving along the 
$z$ axis at the speed of light. 
We refine the MV model by allowing for 
the presence of non-trivial longitudinal correlations between the 
color charges that comprise the nucleons. 
We find that a consistent 
treatment forces us to consider a pancake-like source which moves 
at slightly less than the speed of light.
Our calculation allows us to consider larger values of $\xf$ 
than were allowed in the original MV model. 
\includegraphics{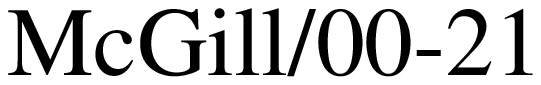}
\end{abstract}


Several years ago, McLerran and Venugopalan realized that
for large enough nuclei at small enough values of the longitudinal
momentum fraction $\xf$, it ought to be possible to compute
the gluon distribution function using 
QCD\cite{MVmodel1}.
Based on this observation, the framework known as the
McLerran-Venugopalan model (MV model) was subsequently
developed\cite{MVmodel2,MVmodel3,MVmodel4,MVmodel9}.
Recently, it was shown\cite{ColorNeut} that the infrared 
divergences present
in the MV model may be cured by capturing the physics of
confinement via a color-neutrality condition to be imposed
on the 
charge density correlation function used as input to the MV model.
In this talk, I will describe 
work\cite{Pancakes} on extending the MV model to larger values of $\xf$.

The MV model as originally formulated in
Refs.~\cite{MVmodel1,MVmodel2,MVmodel3,MVmodel4,MVmodel9},
is restricted to very small longitudinal momentum 
fractions $\xf \lesssim A^{-1/3}/(ma)$, where $A$ is the number
of nucleons, $m$ is the nucleon mass, and $a \sim \LQCD^{-1}$
is the nucleon radius.  In this regime, the longitudinal resolution
of the gluons is so poor that they probe distances which are much
longer than the (Lorentz-contracted) thickness of the nucleus.
Thus, all of the quarks inside the nucleus effectively have the
``same'' value of the longitudinal co\"ordinate $x^{-}$.  At
each value of the transverse co\"ordinate  $\xt$
describing this cr\^epe-shaped
({\it i.e.} very thin)
nucleus, the color charges from a large number of valence quarks
must be summed, resulting in a color charge density which is in a
high-dimensional representation of the gauge group.  This is a necessary
condition for a classical treatment to be valid.   
In going to larger values of $\xf$, we find that the longitudinal
resolution of the gluons improves, and we begin to see the longitudinal
structure of the nucleus.  
In order to include this longitudinal structure,
we are naturally led to a pancake-shaped geometry ({\it i.e.} one
with a finite non-zero thickness).  The details of this fully 
3-dimensional calculation are contained in Ref.~\cite{Pancakes}.

For a classical treatment to be valid, not only must the
color charge be in a large representation of the gauge group, but
the gauge coupling $\alpha_s$ should also be weak.  McLerran
and Venugopalan\cite{MVmodel1} argue that the running coupling 
ought to be evaluated
at the scale $\mu^2$, which is set by the charge-squared per
unit transverse area.  For large enough nuclei, 
$\mu^2 \gg \LQCD^2$, implying that $\alpha_s(\mu^2) \ll 1$.

Thus, we begin our computation of the gluon distribution function
for a large nucleus by solving the {\it classical}\ Yang-Mills 
equations describing a pancake-shaped distribution of color
charge moving along the $z$ axis at nearly the speed of light.
The result is a non-linear expression for the vector 
potential $A(x^{-},\xt)$
in terms of the charge density $\rho(x^{-},\xt)$.    
In the spirit of the Weizs\"acker-Williams
approximation~\cite{WW}, we extract the gluon number density
from the two-point correlation function, 
$\langle A(x^{-},\xt)A(x^{\prime -},\xt') \rangle$.
We replace the quantum mechanical average implied by the
angled brackets with a classical average over an ensemble
of nuclei.  This ensemble is specified by inputting the two-point
charge-density correlator 
$\langle\rho\rho\rangle \sim \DIII(x^{-},\xt)$.
Furthermore, we assume that the correlations are Gaussian.  
Confinement is incorporated into the calculation at this
stage in the form of a color neutrality condition 
on $\DIII$\cite{ColorNeut}.
When $\DIII$ satisfies the color neutrality condition, the
two-point correlation function is infrared finite, and may be
Fourier-transformed to momentum space, producing a gluon number
density $dN/d\xf d^2\qt$ which is differential not only in $\xf$, 
but in the transverse momentum $\qt$ as well.

In the limit $A^{-1/3} \ll 1$, the result of this
rather lengthy calculation reads\footnote{Eq.~(\protect\ref{MasterFormula})
has been written assuming cylindrical geometry for the nucleus.
For a full discussion of the (rather
weak) geometric dependence of the result,
see Ref.~\protect\cite{Pancakes}.}
\beq
\gnumX = 
3AC_F {{2\alpha_s}\over{\pi^2}} \ts
{{1}\over{\xf}} 
\int d^2 \deltat \ts e^{i\qte\cdot\deltate}
\curlyL(\xf,\deltat) \ts
{
{ \exp[N_c\ts \CHI_\infty L(\deltat)] - 1 }
\over
{ N_c \ts\CHI_\infty L(\deltat) }
}.
\label{MasterFormula}
\eeq
Although complicated in appearance, Eq.~(\ref{MasterFormula})
is made up of several easily-understood parts.  The prefactor
shows that at lowest order the number of gluons is simply
proportional to the number of quarks.  
The lowest order result is governed by the function
\beq
\curlyL(\xf,\deltat) \equiv {{1}\over{2}}
\int { {d^2\qt}\over{4\pi^2} } \ts
e^{-i\qte\cdot\deltate} \ts
{
{ \qt^2 \ts \widetilde{\DIII}(\xf,\qt) }
\over
{ [ \qt^2 + (\xf\mn)^2 ]^2 }
},
\label{curlyLdef}
\eeq
and is what would be obtained by considering an Abelian theory.
The non-Abelian corrections to this result are contained in the
exponential factor, which depends on two quantities.  First, the
the spatial dependence is determined by
\beq
L(\deltat)   \equiv
\int { {d^2\qt}\over{4\pi^2} } \ts
{
{ \widetilde{\DIII}(0,\qt) }
\over
{  \qt^4  }
}
\ts \Bigl[
e^{-i\qte\cdot\deltate}-1
\Bigr] .
\label{plainLdef}
\eeq
The strength of the non-Abelian corrections is set by
the factor
\beq
\CHI_\infty =
{{1}\over{\pi R^2}}
{{3Ag^4 C_F}\over{N_c^2-1}}
\sim
8\pi \alpha_s^2 A^{1/3} \LQCD^2.
\label{CHIdef}
\eeq
These effects are most prominent in very large nuclei.
For uranium we have 
$\CHI_\infty \sim 5$ or $6 \LQCD^2$.
To obtain $\CHI_\infty = 20\LQCD^2$ (as is employed in the plots
below) requires of order $10^4$ nucleons.
Finally, we note that 
if we set $\xf = 0$ in Eqs.~(\ref{MasterFormula})--(\ref{plainLdef}), 
we explicitly 
reproduce the original MV result~\cite{MVmodel9}.

The functions appearing in Eqs.~(\ref{curlyLdef}) and~(\ref{plainLdef})
are not finite at $\qt=0$ unless the charge density correlator 
$\DIII$ satisfies the requirement of color neutrality.
As explained in Ref.~\cite{ColorNeut}, a key consequence of confinement
is the appearance of color neutral nucleons.  Mathematically,
this consequence may be implemented as a constraint on $\curlyD$:
\beq
\int d\Delta^{-} d^2\deltat \ts\ts \DIII(\Delta^{-},\deltat) = 0,
\qquad\hbox{or}\qquad \widetilde\DIII(0,{\bf 0}) = 0.
\label{Key}
\eeq
Eq.~(\ref{Key}) ensures that widely-separated
nucleons are essentially uncorrelated.  In order for~(\ref{Key})
to be true, $\curlyD$ must contain a length scale.  Not surprisingly,
this scale turns out to be the nucleon radius $a\sim \LQCD^{-1}$.
When combined with the assumption of rotational
symmetry in the transverse plane, Eq.~(\ref{Key})
is sufficient to render the
functions contributing to the gluon number density completely 
infrared finite:  the integrals get cut off at 
$q \sim a^{-1} \sim \LQCD$.

Aside from the neutrality condition~(\ref{Key}), the correlation
function $\DIII$ is unspecified.  In order to illustrate the
general features of the gluon number density~(\ref{MasterFormula}),
it is convenient to choose the following form for $\widetilde\DIII$:
\beq
\widetilde\DIII(\xf,\qt) =
1 - 
{ { 1 } \over  {  1 + (a\qt)^2 + (\xf\mn a)^2 }  }.
\label{monopole}
\eeq
In the context of Kovchegov's nuclear model\cite{Model},
the correlation function given in Eq.~(\ref{monopole})
corresponds to quarks which are distributed within the
nucleons according to the weight
\beq
\vert \psi(\vec{r}\ts)\vert^2 =
{ {1}\over{2\pi^2a^3} } \ts
{{a}\over{r}}
\ts K_1\biggl({{r}\over{a}}\biggr).
\label{qdist}
\eeq
At large distances, the modified Bessel function produces
a Yukawa-like behavior in this probability distribution.


\begin{figure}[b] 
\includegraphics{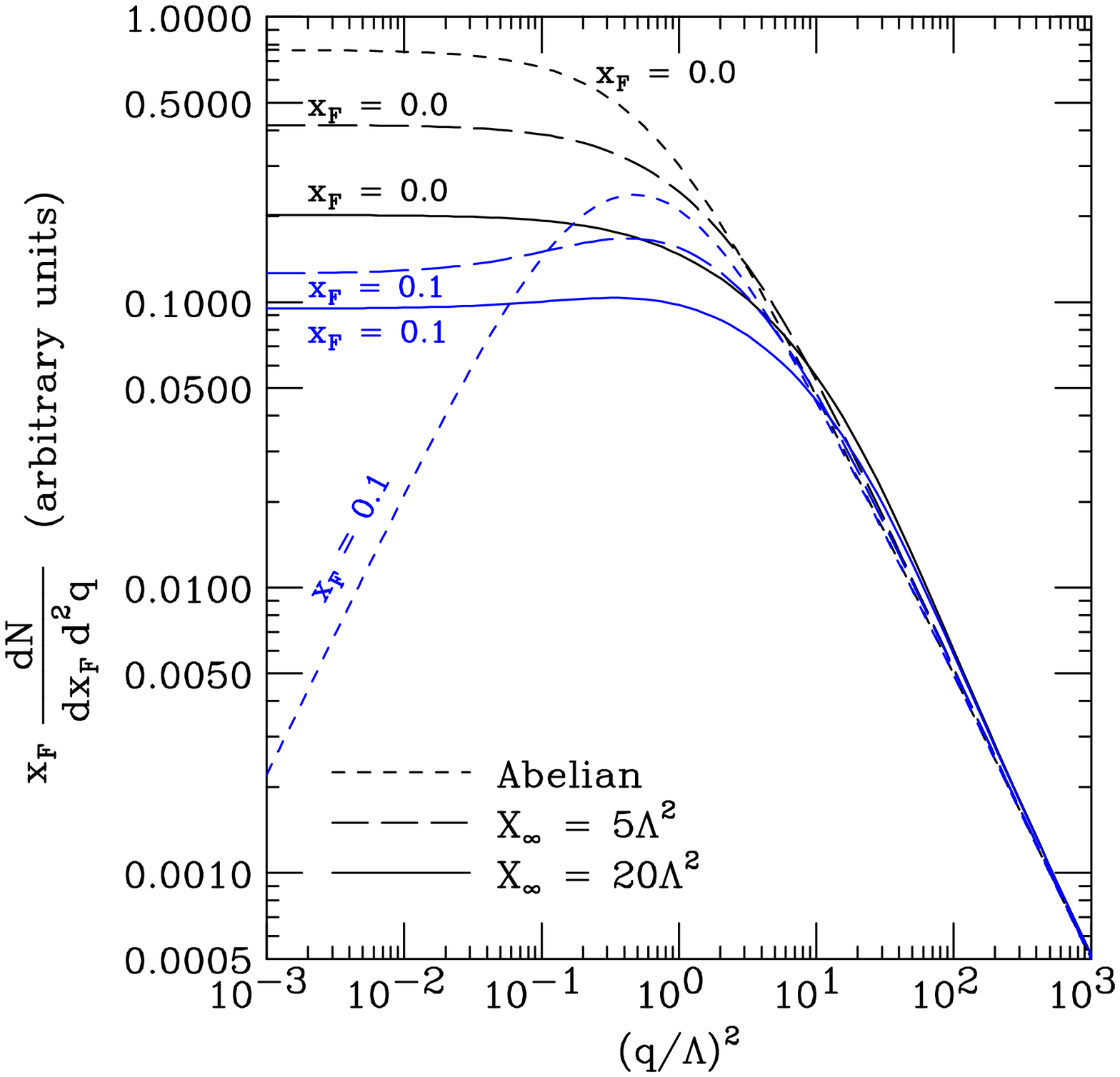}
\includegraphics{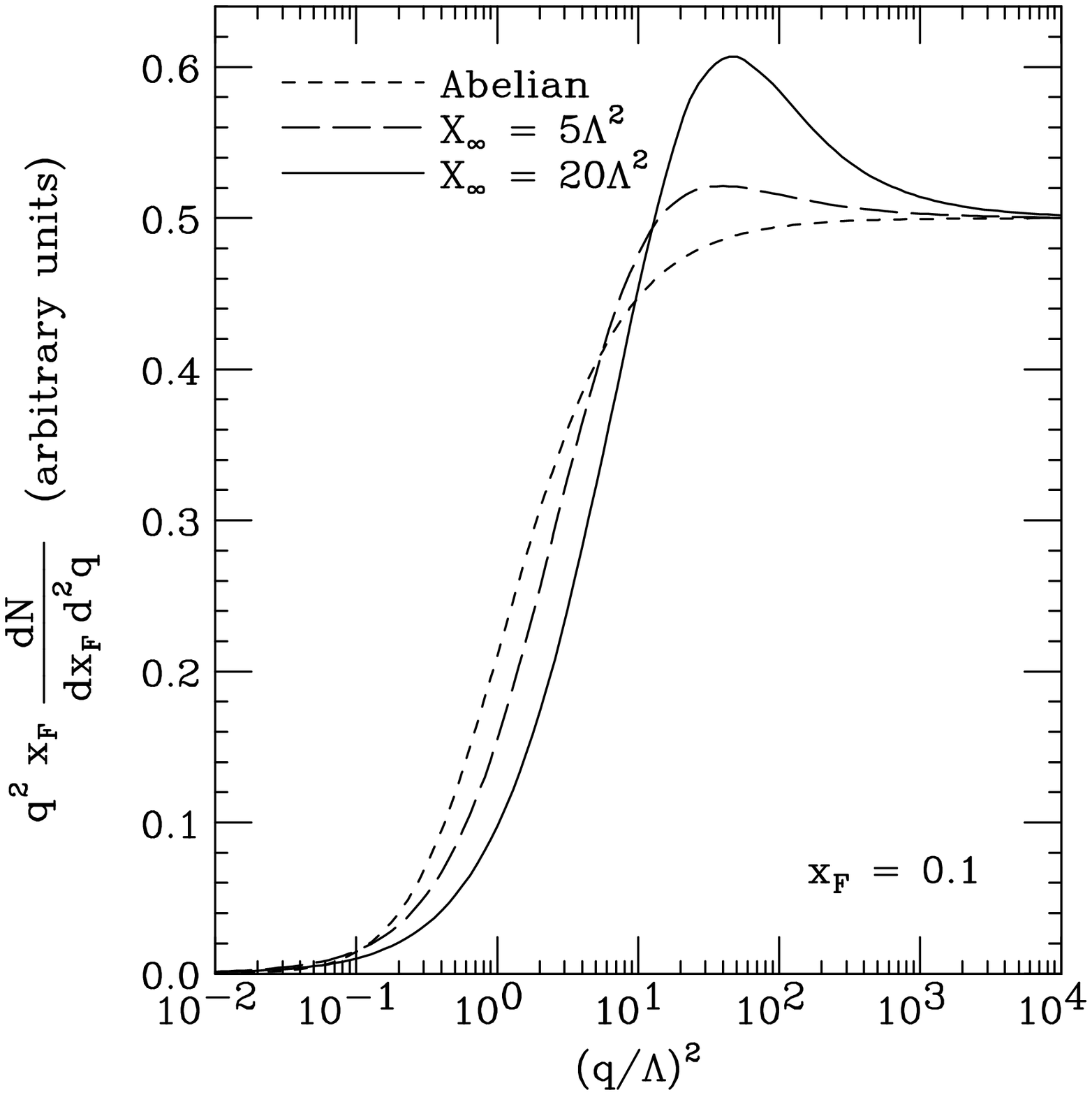}
\vspace{2.7in}
\begin{minipage}[b]{2.8in}
\caption{Fully differential gluon number density 
plotted versus $\qqq^2$ at $\xf = 0.0$ and $0.1$.
The three curves are for
$\CHI_\infty = 0$ (Abelian), $5\LQCD^2$, and $20\LQCD^2$.
The gluon density saturates at small $\qqq^2$.
}
\label{FullDiff}
\end{minipage}
\hfill
\begin{minipage}[b]{2.8in}
\caption{Fully differential gluon number density 
at $\xf = 0.1$ multiplied by $\qqq^2$ versus $\qqq^2$ for 
$\CHI_\infty = 0$ (Abelian), $5\LQCD^2$, and $20\LQCD^2$.
The area under each curve accurately reflects the contributions
to $g_A(\xf,Q^2)$.
}
\label{AreaPlot}
\end{minipage}
\end{figure}

Fig.~\ref{FullDiff} exhibits the behavior of the fully differential
gluon number density as a function of the transverse momentum $\qt^2$.
We see that there is saturation:  the number of gluons increases
as $\qt^2 \rightarrow 0$ to some maximum value and then stops
growing.  Qualitatively, these distributions are very much like
those obtained by Muller from the point of view of onium 
scattering\cite{Muller1,Muller2}.

The gluon distribution function resolved at the scale $Q^2$
is related to the fully differential gluon number density by
\beq
g_A(\xf,Q^2)  \equiv 
\int_{\vert\qte\vert\le Q}\nts\nts  d^2\qt \ts
{{dN}\over{d\xf d^2\qt}}.
\label{PDF}
\eeq
In Fig.~\ref{AreaPlot} we multiply
the fully differential distribution by $\qt^2$: on the semi-log
scale used this produces a ``true'' representation of where
the important contributions to Eq.~(\ref{PDF}) are.   We see
that the very low $\qt^2$ region does not play a significant role
provided $Q^2$ is not too small.  From the plot we see that for
$Q^2\rightarrow\infty$, the gluon distribution function becomes
insensitive to the presence or absence of the non-Abelian corrections.
In fact, we can prove under rather general circumstances that
\beq
\int d^2\qt \ts
\Biggl\{ 
\gnumX\Biggl\vert_{{\rm all}\atop{\rm orders}} \nts\nts
-  \ts \gnumX\Biggl\vert_{{\rm lowest}\atop{\rm order}} 
\Biggr\}
= 0,
\eeq
independent of $\DIII$~\cite{ColorNeut}.  

\begin{figure}[h] 
\includegraphics{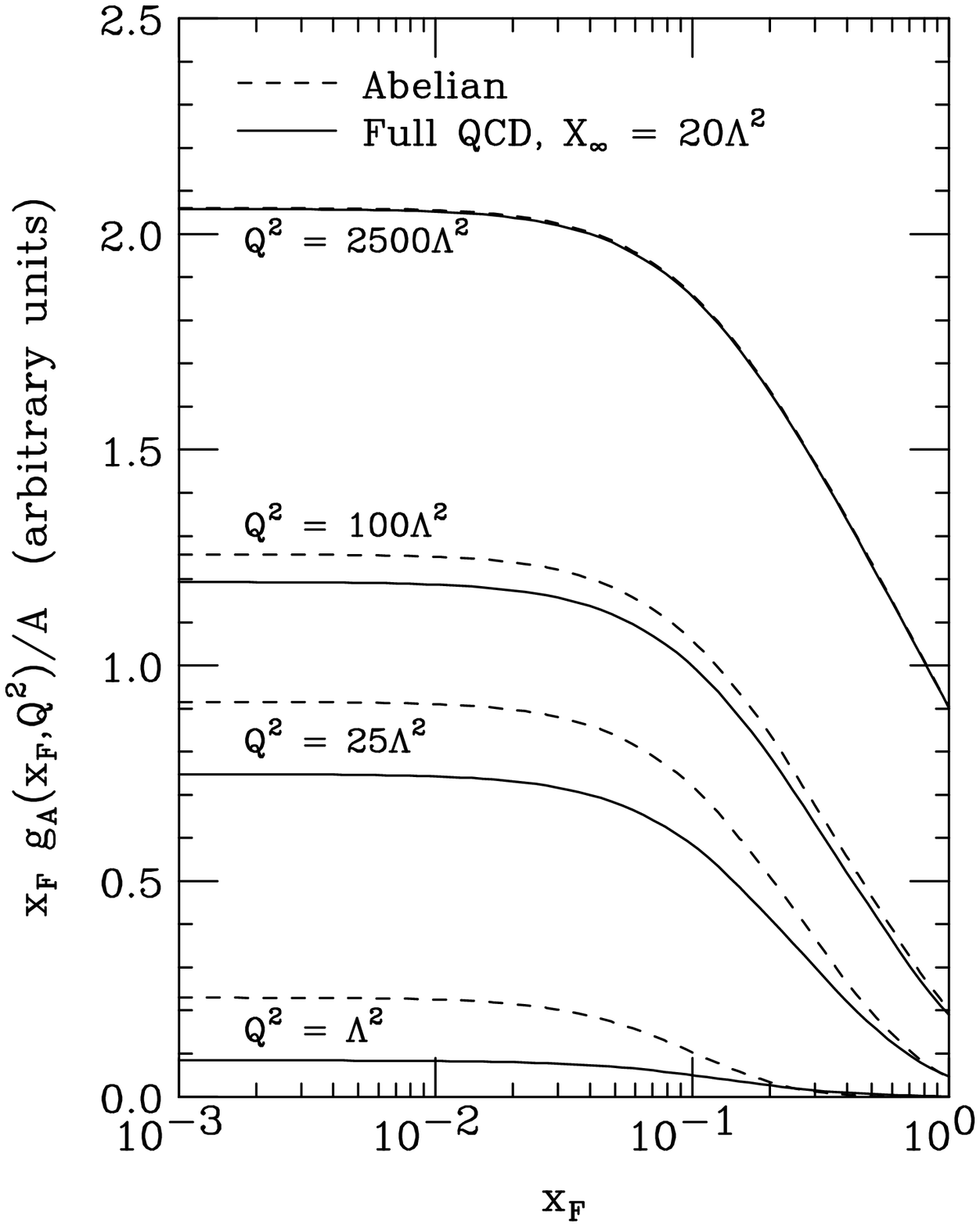}
\includegraphics{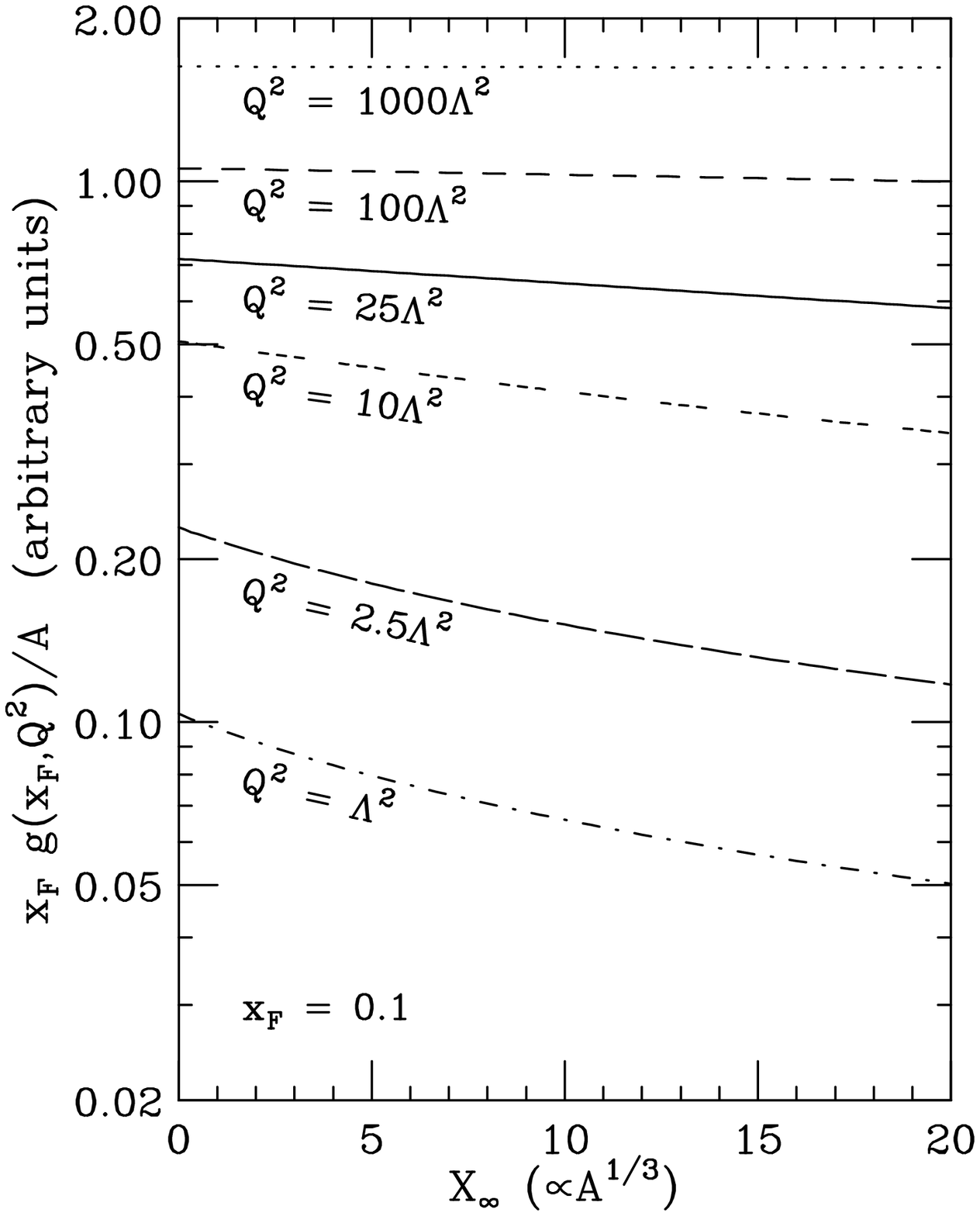}
\vspace{3.3in}
\begin{minipage}[b]{2.8in}
\caption{The gluon structure function $\xf g_{{}_A}(\xf,Q^2)$
plotted versus $\xf$ for several values of $Q^2$.
The dashed curves illustrate the effect of ignoring the
non-Abelian corrections, which are included in the solid curves.
}
\label{g-plot}
\end{minipage}
\hfill
\begin{minipage}[b]{2.8in}
\caption{Nuclear dependence of the gluon structure function
$\xf g_{{}_A}(\xf,Q^2)$.  For sufficiently small $Q^2$
the number of gluons grows more slowly than the number of
nucleons as $A$ is increased.
}
\label{NucDep}
\end{minipage}
\end{figure}

In Fig.~\ref{g-plot}, we exhibit
$\xf g_A(\xf,Q^2)$ as a function of $\xf$ for several
different values of $Q^2$.  At low values of $Q^2$,
the addition of the non-Abelian corrections reduces the 
number of gluons from the Abelian result, while by the
time $Q^2 = 2500 \LQCD^2$ is reached, the effect of the
non-Abelian terms is negligible.  Although we have plotted
out to $\xf=1$, the calculation is not reliable
beyond $\xf\sim 0.25$\cite{Pancakes}.  For small $\xf$ the
gluon distribution function exhibits the pure $1/\xf$ dependence
characteristic of the original MV model.

Finally, we present Fig.~\ref{NucDep}, which illustrates
the deviation of the gluon distribution function from the
na{\"\i}ve expectation that for $A$ nucleons we should obtain
$A$ times the result for a single nucleon.  We see that at
low values of $Q^2$ the distribution function grows less
rapidly than the number of nucleons as $A$ is increased,
whereas at large $Q^2$, the simple scaling expectation holds.


\begin{references}
 
\bibitem{MVmodel1}
L.~McLerran and R.~Venugopalan,
{\it Phys. Rev.} {\bf D49}, 2233 (1994). [hep-ph/9309289]

\bibitem{MVmodel2}
L.~McLerran and R.~Venugopalan,
{\it Phys. Rev.} {\bf D49}, 3352 (1994). [hep-ph/9311205]

\bibitem{MVmodel3}
L.~McLerran and R.~Venugopalan,
{\it Phys. Rev.} {\bf D50}, 2225 (1994). [hep-ph/9402335]

\bibitem{MVmodel4}
A.~Ayala, J.~Jalilian-Marian, L.~McLerran,
and R.~Venugopalan,
{\it Phys. Rev.} {\bf D52}, 2935 (1995). [hep-ph/9501324]

\bibitem{MVmodel9}
J.~Jalilian-Marian, A.~Kovner, L.~McLerran, and H.~Weigert,
{\it Phys. Rev.} {\bf D55}, 5414 (1997). [hep-ph/9606337]

\bibitem{ColorNeut}
C.S.~Lam and G.~Mahlon, 
{\it Phys. Rev.} {\bf D61}, 014005 (2000). [hep-ph/9907281]

\bibitem{Pancakes}
C.S.~Lam and G.~Mahlon, 
``Longitudinal Resolution in a Large Relativistic Nucleus:
 Adding a Dimension to the McLerran-Venugopalan Model,''
hep-ph/0007133.

\bibitem{WW}
C.~Weizs\"acker and E.~Williams,
{\it Z. Phys.} {\bf 88}, 244 (1934).

\bibitem{Model}
Yu. Kovchegov,
{\it Phys. Rev.} {\bf D54}, 5463 (1996). [hep-ph/9605446]

\bibitem{Muller1}
A.~Mueller, 
{\it Nucl. Phys.} {\bf B335}, 115 (1990).

\bibitem{Muller2}
A.~Mueller,
{\it Nucl. Phys.} {\bf B558}, 285 (1999). [hep-ph/9904404]



\end{references}
\end{document}